\documentclass[10pt,table,xcdraw]{article} 
\usepackage[preprint]{tmlr}

\usepackage{booktabs}
\usepackage{multicol, multirow}
\usepackage{amsmath}
\usepackage{amssymb}
\usepackage{mathtools}
\usepackage{amsthm}
\usepackage{float}
\usepackage{graphicx}
\usepackage{algorithm}
\usepackage{algorithmic}

\usepackage{amsmath,amsfonts,bm}

\def\eqref#1{equation~\ref{#1}}

\def\1{\bm{1}}

\DeclareMathAlphabet{\mathsfit}{\encodingdefault}{\sfdefault}{m}{sl}
\SetMathAlphabet{\mathsfit}{bold}{\encodingdefault}{\sfdefault}{bx}{n}

\usepackage{hyperref}

\let\classAND\AND
\let\AND\relax

\let\AND\classAND

\AtBeginEnvironment{algorithmic}{\let\AND\algoAND}

\title{Interpretable CNN-Multilevel Attention Transformer for Rapid Recognition of Pneumonia from Chest X-Ray Images}

\author{\name Shengchao Chen \email pavelchen@ieee.org \\
      \addr School of Information and Communication Engineering, Hainan University
      \AND
      \name Sufen Ren \email rensufen@hainanu.edu.cn \\
      \addr School of Information and Communication Engineering, Hainan University
      \AND
      \name Guanjun Wang \email wangguanjun@hainanu.edu.cn \\
      \addr School of Information and Communication Engineering, Hainan University
      \AND
      \name Mengxing Huang \email huangmx09@hainanu.edu.cn \\
      \addr School of Information and Communication Engineering, Hainan University
      \AND
      \name Chenyang Xue \email xuechenyang@nuc.edu.cn \\ 
      \addr School of Instrument and Electronics, North University of China}

\def\month{MM}  
\def\year{YYYY} 
\def\openreview{\url{https://openreview.net/forum?id=XXXX}} 

\begin{document}

\maketitle

\textcolor{red}{\textbf{This paper has been accepted by the IEEE Journal of Biomedical and Health Informatics (IEEE JBHI). More information is available at \hyperlink{10.1109/JBHI.2023.3247949}{this website/doi}.}}
\begin{abstract}
Chest imaging plays an essential role in diagnosing and predicting patients with COVID-19 with evidence of worsening respiratory status. Many deep learning-based approaches for pneumonia recognition have been developed to enable computer-aided diagnosis. However, the long training and inference time makes them inflexible, and the lack of interpretability reduces their credibility in clinical medical practice. This paper aims to develop a pneumonia recognition framework with interpretability, which can understand the complex relationship between lung features and related diseases in chest X-ray (CXR) images to provide high-speed analytics support for medical practice. To reduce the computational complexity to accelerate the recognition process, a novel multi-level self-attention mechanism within Transformer has been proposed to accelerate convergence and emphasize the task-related feature regions. Moreover, a practical CXR image data augmentation has been adopted to address the scarcity of medical image data problems to boost the model's performance. The effectiveness of the proposed method has been demonstrated on the classic COVID-19 recognition task using the widespread pneumonia CXR image dataset. In addition, abundant ablation experiments validate the effectiveness and necessity of all of the components of the proposed method.
\end{abstract}

\section{Introduction}
COVID-19 was first discovered in December 2019 in Wuhan, Hubei Province, China, and its ability to spread robustly has led to a global pandemic of it and its variant strains (Omicron BA.1, Omicron BA.2, etc.)~\cite{cucinotta2020declares}. As of September 10, 2022, more than 607,362,097 people have been diagnosed with COVID-19 positive worldwide, with a death toll of 6,510,110. Its main symptoms include fever, shortness of breath, and loss of taste~\cite{lauer2020incubation}. In some cases, the infection can lead to serious respiratory complications, most notably pneumonia, which often requires admission to Intensive Care Unit (ICU). Unfortunately, the dramatic increase in diseases has paralyzed many health systems, and the shortage of wards and medical equipment has forced many countries to designate a series of epidemic prevention policies, such as distance restraints and national lockdowns~\cite{phua2020intensive}.

Real-time reverse transcription polymerase chain reaction (RT-PCR) is the primary diagnostic tool for COVID-19~\cite{ieracitano2022fuzzy}. In clinical practice, chest X-ray (CXR) have long been used to evaluate patients' lungs because of their portability and cost-effectiveness compared to computed tomography (CT)~\cite{wong2020frequency}. However, the resulting images require manual analysis by radiologists, making the diagnostic process inefficient and difficult to cater to large-scale diagnostic needs. This diagnostic strategy, which requires the practical intervention of experts and is time-consuming and laborious, needs urgent improvement.

Rapid advances in deep learning (DL) techniques have enabled the development of a practical computer-aided diagnostic tool to detect COVID-19-related diseases from large-scale CXR images. DL allows computational models consisting of multiple layers that contain several processing units to learn distinct feature abstraction, thus avoiding direct interaction with skilled experts. For example, lots of work trains convolutional neural network (CNNs)-based models to recognize COVID-19 from CXR images~\cite{wang2020covid, heidari2020improving, apostolopoulos2020covid, islam2020combined, karthik2021learning, khan2020coronet, ezzat2021optimized, umer2022covinet}, the key is to establish the relationship between the feature distribution in lung region and disease labels. However, the shape and location of lungs, speckled, hazy, irregular, hairy glassy cloudiness, and reticular cloudiness~\cite{jacobi2020portable} contribute to the recognition results to varying degrees, which means not all features are efficient and discovering task-related features is crucial to boost the recognition accuracy. Inspired by the human visual system, the attention mechanism has been widely used in image-based tasks~\cite{dosovitskiy2020image, mnih2014recurrent, wang2018non, hu2018squeeze, fu2019dual, yuan2021ocnet, huang2019ccnet}. These encourage models to focus on task-related information regions through reshaping features while suppressing the interference of useless peripheral information to boost performance. Nevertheless, adding the attention mechanism inevitably increases the computational complexity, indirectly leading to its slowing down of learning and inference, contrary to the original intention of achieving high-speed recognition.

What's worse, the superiority of DL relies heavily on pre-collected datasets, a common bottleneck of data-driven approaches. The major problem with medical images is the disadvantage in size due to the cumbersome collection process, making it challenging to train a robust network adequately. Image-based augmentation approaches, such smoothing, noise addition, flipping, image masking, are considered an effective solution to above problem by extending datasets. However, their generations' distribution may be detached from the original image distribution. DL-based augemntation methods such as autoencoders (AEs)~\cite{ng2011sparse} and generative adversarial networks (GANs)~\cite{goodfellow2020generative} are widely used to generate data with similar distribution to the original dataset~\cite{mariani2018bagan, hsu2017unsupervised}. Unfortunately, requiring tedious data processing and network training makes them hard to satisfy the low-resourced medical applications . Moreover, DL models that regarded as 'black boxes' because of large randomness are hard to gain confidence in practical CXR images-based pneumonia recognition tasks~\cite{liu2022deep}.

Therefore, this paper aims to design a framework with interpretability and fast recognition of pneumonia, especially COVID-19-positive, for providing reliable analytic support to computer-aided medical diagnosing systems.

To this end, we propose a model cascades CNN and Transformer, dubbed CMT, to understand complex lung's feature correlation with specific pneumonia with CXR images. First, pre-trained CNN-based models have been considered feature extractors for abstracting complex features from input CXR images and covert them into feature embedding. Secondly, the Transformer gets these embedding to explore the importance of a feature so that we understand the relationship between task-related features and specific pneumonia diseases.

Specifically, to relieve the negative implication of medical data scarcity on recognition performance, a practical CXR image data augmentation is proposed to extend the dataset as the pre-work of recognition, which generated new images undeviating from original distribution at a low cost. Moreover, we introduce a novel self-attention mechanism within Transformer, named \textbf{M}ultilevel \textbf{M}ulti-Head \textbf{S}elf-\textbf{A}ttention (MMSA), to reduce the overall computational complexity while keeping excellent recognition accuracy by computing different levels' importance. The proposed CMT provides a novel platform for pneumonia recognition and can be extended to other medical image recognition tasks.

To verify the effectiveness of the proposed CMT, We fused patient lung images collected from five different medical institutions to produce a complete pneumonia identification dataset and conducted extensive experiments on this dataset. The main contribution of this work can be summarized in four-fold: 
\begin{itemize}
    \item To provide high-efficient pneumonia (especially COVID-19) recognition solution with credibility in a clinic medical diagnosing practices, we proposed \textbf{C}NN-\textbf{M}MSA-\textbf{T}ransformer (CMT) that cascaded CNN and Transformer with low computational complexity attention mechanism, to analyze the CXR images.

    \item To reduce the computational complexity of Transformer and improve the task-related lung feature perception ability for adapting low-resourced medical applications,  we propose \textbf{M}ultilevel \textbf{M}ulti-Head \textbf{S}elf-\textbf{A}ttention (MMSA) \textit{\textbf{(Section \ref{MMSA})}}, which outperform conventional attention mechanisms with ultra-fast convergence and powerful feature perception while guaranteeing excellent performance.
    
    \item To mitigate the negative impact of medical image data scarcity on the model recognition effect, a random beta sampling (RBS)-based feature fusion data augmentation approach, named FFA \textit{\textbf{(Section \ref{FFA})}}, was proposed by distributing non-divergent augmentation of the original CXR image dataset.
    
    \item To boost the credibility of the proposed CMT in clinic medical diagnosing applications, the model's interpretability has been exlpored and validated by by visulizing the weight matrix and feature activation map during inference process \textit{\textbf{(Section \ref{INTER})}}.

\end{itemize}

The remainder of this work is in the following: Section \ref{RW} reviews related works. Section \ref{MH} describes the proposed approach in detail. Section \ref{EXP} describes the specific implementation details of the experiment and shows the results of experiments. Section \ref{CF} concludes this work.

\section{Related Work}
\label{RW}
This section summarizes the relevant representative approaches on this work, which are divided into three parts, including COVID-19 recognition with CNN, Self-Attention mechanism, and Vision Transformers (ViTs).

\subsection{COVID-19 Recognition with CNN}
Since COVID-19 has become widespread, many works have described the application of CN and DL in diagnosing the disease based on medical images and have achieved promising results. CNNs such as ResNet\cite{he2016deep}, VGGNet\cite{simonyan2014very}, and AlexNet\cite{krizhevsky2012imagenet} have shown excellent performance in computer vision tasks such as object detection, image classification, and semantic segmentation. Lots of works have recently applied the CNN-based models in COVID-19 recognition~\cite{wang2020covid,heidari2020improving,apostolopoulos2020covid,islam2020combined,khan2020coronet,karthik2021learning,khan2020coronet,ezzat2021optimized,umer2022covinet}. For example, Wang \textit{et al.} proposed COVID-Net to perform a three-way recognition in COVID-19/Non-COVID-19~\cite{wang2020covid}. Heidari \textit{et al.} used vanilla VGGNet to classify COVID-19-based disease from CXR image~\cite{heidari2020improving}. Considering the low-resourced computer constrain in medical institutions, fully training a CNN-based model based on knowledge is expensive. For this reason, Apostolopoulos \textit{et al.} applied a pre-trained CNN-based model based on transfer learning (TL) to achieve three-way CXR image recognition while satisfying the computational restriction~\cite{apostolopoulos2020covid}. A series of related works that used pre-trained models to accelerate the convergence has surface, e.g., CNN-LSTM~\cite{islam2020combined}, CoroNet~\cite{khan2020coronet}, etc. However, these methods ignore other pneumonia that remains in the CXR image dataset that not limited to classifying COVID-19/Non-COVID-19, so that could be lead to a high cost to build different models to recognize different pneumonia. To overcome the problem, some researchers view the COVID-19 recognition task as a multi-class classification problem~\cite{sun2021mfbcnnc,bhowal2021choquet,mondal2021xvitcos,nandal2021fuzzy,kumar2022sars,malhotra2022multi,wu2021ulnet,esmi2023fuzzy}. Among them, the study~\cite{sun2021mfbcnnc} introduced an optimized biogeography-based biogeography network and a momentum factor biogeography CNN to detect COVID-19 without handmade models' hyperparameters. An ensemble deep learning model based on coalition-game theory, information theory, and Lambda fuzzy approximation method is proposed to detect three different types of pulmonary symptoms in CXR images~\cite{bhowal2021choquet}. Kumar \textit{et al.} introduced a CADx system combining graph convolutional network (GCN) and CNN, named SARS-Net, for classification and detection of abnormalities in COVID-19 infection~\cite{kumar2022sars}.

However, the above studies have three primary limitations: \textbf{i)} viewing the COVID-19 recognition as a binary classification problem lack flexibility in practical diagnosis scenarios; \textbf{ii)} achieving multi-symptom recognition by building multiple paths or stacking multiple binary classifiers is an expensive strategy; \textbf{iii)} some relies on a series of feature engineering that based on skilled experts to extract critical information so that against the end-to-end principle.

\subsection{Self-Attention Mechanism}
Humans tend to focus more on purpose-related content in life and ignore irrelevant information. Self-attention mechanisms, inspired by the human visual system, are often used to enhance the essential information and suppress the remaining invalid noisy details\cite{dosovitskiy2020image,mnih2014recurrent,wang2018non,fu2019dual,yuan2021ocnet,guo2022beyond,wu2022pale,shen2020global,liu2021swin}. Mni \textit{et al.} \cite{mnih2014recurrent} combined recurrent neural networks and attention mechanisms to obtain excellent image classification performance. Wang \textit{et al.} \cite{wang2018non} used self-attention to capture feature relevance based on residual networks. Hu \textit{et al.} \cite{hu2018squeeze} applied inter-channel attention to model the interdependencies between channels to recalibrate channel feature responses adaptively. Fu \textit{et al.} \cite{fu2019dual} applies two different parallel self-attention aggregations to semantic segmentation. Yuan \textit{et al.} \cite{yuan2021ocnet} uses a self-attention with ASPP to exploit contextual dependencies. Huang \cite{huang2019ccnet} applies Criss-Cross attention to capture global relevance information. In summary, self-attention is often used to learn feature representations by recalibrating the feature graph for features. However, the high computational complexity is a serious problem within these attention mechanisms. These studies~\cite {wu2022pale, shen2020global,liu2021swin} consider the potential drawback that the high computational complexity brings by capturing global attention to executing self-attention in relatively more minor local regions. Based on the above problem and research foundation, this paper aims to design a novel CXR image recognition framework to be applied in practical COVID-19 diagnosis applications. The core is a cost-effective self-attention mechanism that can learn the comprehensive representation of global-local features while keeping low computational and architectural complexity.

\subsection{Vision Transformers (ViTs)}
Transformer \cite{vaswani2017attention} relies entirely on self-attention to deal with long-term dependencies in sequential data. Inspired by the outstanding results of Transformer for natural language processing (NLP)\cite{devlin2018bert, dai2019transformer}, Vision Transformers (ViTs) \cite{dosovitskiy2020image} were proposed for encoding image token sequences. To improve the applicability and performance of transformers in image data processing, Dosovitskiy \textit{et al.}\cite{dosovitskiy2020image} exploited the proposed use of embedded patch segmentation to partition the original image into several small patches and make tokens for them separately, aiming to represent the local structure through the relationships between patches. Touvron \textit{et al.} \cite{touvron2021going} built deep network models by improving the attention mechanism, i.e., introducing formaldehyde and class-specific attention for each channel. Fan \textit{et al.} \cite{fan2021multiscale} proposed a multiscale ViT structure with multiscale feature pyramids. Zhou \textit{et al.} \cite{zhou2021deepvit} increased their diversity at different levels of representation by regenerating attention maps. However, these methods inevitably bring higher computational cost while enhancing performance. This paper proposes a pneumonia recognition framework based on Transformer with a novel low computational complexity attention mechanism, which can effectively capture the relationship between global and local features while maintaining the high-speed training and inference process.

\section{Methodology}
\label{MH}
\begin{figure*}[htp]
    \centering
    \includegraphics[width=0.75\textwidth]{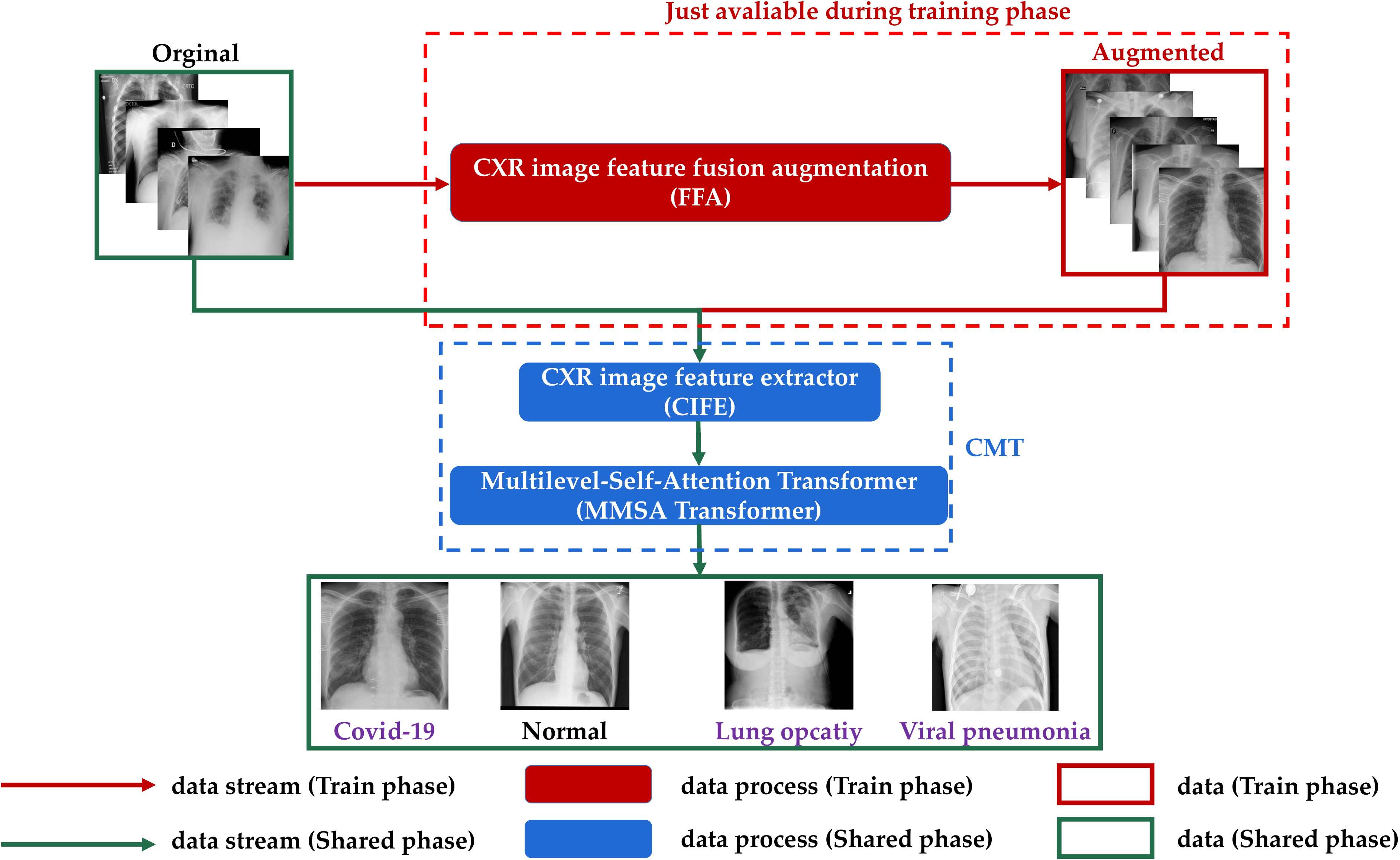}
    \caption{Schematic diagram of the proposed CMT, which mainly consists of a CXR image feature fusion augmentation (FFA) module, CXR image feature extractor (CIFE), and Multilevel Multi-Head Self-Attention Transformer (MMSA Transformer). Note that the Shared phase means the training and inference phase.}
    \label{F1}
\end{figure*}

The architecture of the proposed recognition framework is shown in Fig. \ref{F1}, which consists of two parts. The first part is the CXR image feature fusion data augmentation (FFA) that takes effect only in the model training phase, which receives the original CXR images and generates augmented images with a similar distribution. The second part is the CMT consisting of a CXR image feature extractor (CIFE) and Multilevel-Self-Attention Transformer (MMSA Transformer), where CIFE is used to extract deeper feature information in CXR images. MMSA Transformer is used to receive this feature information to explore and construct complex nonlinear relationships between global feature, local feature, and disease information. The CMT receives data from the original and augmented datasets for training. The details of their design will be specified next.

\subsection{CXR Image Feature Fusion Augmentation (FFA)}
\label{FFA}
Unlike other types of image dataset, the scarcity of medical image data hinders further performance improvement of data-driven approaches. An efficient CXR image augmentation strategy, named \textbf{F}eature \textbf{F}usion \textbf{A}ugmentation (FFA), is proposed to overcome this problem. FFA is used for online augmentation based on the original CXR images to extend the original dataset to improve recognition accuracy. The specific implementation of the proposed FFA algorithm is shown in Algorithm 1. The basic idea is to randomly fuse two CXR images under the same category based on the weight matrix generated by random beta sampling (RBS), which makes the generated images not deviate from the original distribution while maintaining the essential features.

The schematic diagram of the FFA algorithm is shown in Fig.~\ref{FFA}. Specifically, first, two CXR images, $\mathbf{I_1} \in \mathbb{R}^{h\times w}$ and $\mathbf{I_2} \in \mathbb{R}^{h\times w}$, under the same class are randomly selected and divided equally according to patch size $p$, such that the size of the divided region is $\mathbf{I_{d}} \in \mathbb{R}^{\frac{h}{p} \times \frac{w}{p}}$. Then, fixed weights $z'$ and $1-z'$ are obtained according to RBS, and similarly, RBS is applied at each point of the $\frac{h}{p} \times \frac{w}{p}$ region to obtain different weights $\hat{z}$ and $1-\hat{z}$. The Beta distortion is the probability distribution over the continuous variable $x \in [0,1]$, which can be formulated as Eq.(1). 

\begin{equation}
\begin{aligned}
\operatorname{Beta}(\mathrm{x} \mid \mathrm{a}, \mathrm{b})=\frac{\Gamma(\mathrm{a}+\mathrm{b})}{\Gamma(\mathrm{a}) \Gamma(\mathrm{b})} \mu^{\mathrm{a}-1}(1-\mu)^{\mathrm{b}-1} \\=\frac{1}{\mathrm{~B}(\mathrm{a}, \mathrm{b})} \mu^{\mathrm{a}-1}(1-\mu)^{\mathrm{b}-1}
\end{aligned}
\end{equation}
where $a$ and $b$ are determined parameters that are constants and more than 0, and $\Gamma(x)$ and $B(x, y)$ is Gamma function and Beta function that can be formulated as Eq.(2) and Eq.(3).
\begin{equation}
\Gamma(\mathrm{x})=\int_0^{+\infty} \mathrm{t}^{\mathrm{x}-1} \mathrm{e}^{-\mathrm{t}} \mathrm{dt},
\end{equation}
\begin{equation}
B(a, b)=\int_0^1 t^{a-1}(1-t)^{b-1} \mathrm{dt}.
\end{equation}
RBS, on the other hand, randomly selects one of the beta divisions calculated from the pre-determined $a$, $b$ as the weights. These weights are combined as the weight matrices $\mathbf{M} \in \mathbb{R}^{h \times w}$ and $\mathbf{M_d} \in \mathbb{R}^{h \times w}$, respectively. It is worth noting that the weights of the points in the $\frac{h}{p} \times \frac{w}{p}$ region of the matrix are inconsistent. In contrast, the other regions' weights are consistent, allowing the generated images to ensure the presence of the necessary features. Finally, the two images and the weight matrices are Hadamard-product and summed, as shown in Line 11 of Algorithm 1, to obtain the generated image $\mathbf{\hat{I}} \in \mathbb{R}^{h \times w}$.

\begin{algorithm}[htp]
\caption{CXR image feature fusion augmentation (FFA) algorithm.}\label{alg:alg1}
\begin{algorithmic}
\STATE $\mathbf{Input:}$ Two random CXR Images matrix $\mathbf{I_1}$, $\mathbf{I_2}$, pacth size $p$, random coefficient $\alpha$ less than 1, and positional parameter $m$, $n$ for the feature fusion augmentation system $FFA = (\mathbf{I_1}, \mathbf{I_2}, p, \alpha, m, n)$.
\STATE $\mathbf{Output:}$ Generated image $\mathbf{\hat{I}}$;
\STATE 1: $z'$ = $BRS(\alpha = \alpha,\hspace{0.1cm}\beta = \alpha)$; $z_{d}'=1 - z'$; \\// Step 1: Random Beta Sampling (RBS) to generate fixed weights $z'$ and $z_{d}'$
\STATE 2: Define two all-zero matrix $\mathbf{M}$ and $\mathbf{M_d}$; 
\STATE 3: Define the $h$ and $w$ denotes the height and weight of the image matrix $\mathbf{I_1}$ and $\mathbf{I_2}$.
\STATE 4: $\mathbf{for}$ $(i=0;\hspace{0.1cm}i \leq h;\hspace{0.1cm}i++)$ $\mathbf{do}$
\STATE 5: \qquad $\mathbf{for}$ $(j=0;\hspace{0.1cm}j \leq w;\hspace{0.1cm}j++)$ $\mathbf{do}$
\STATE 6: \qquad\qquad $\mathbf{if}$ $m * \frac{h}{p}$ \textless $i$ \textless $(m+1) * \frac{h}{p}$ and $n * \frac{w}{p}$ \textless $j$ \textless $(n+1) * \frac{w}{p}$ $\mathbf{then}$
\STATE 7: \qquad\qquad\qquad $\hat{z}$ = $BRS(\alpha = \alpha,\hspace{0.1cm}\beta = \alpha)$; $\hat{z}'=1 - z'$;\\ \qquad\qquad\qquad\quad// Step 2: Generate random weights using random beta sampling in the loop
\STATE 8: \qquad\qquad\qquad $\mathbf{M}\hspace{0.1cm}[i,j] = \hat{z}$; $\mathbf{M_d}\hspace{0.1cm}[i,j] = \hat{z}'$
\STATE 9: \qquad\qquad $\mathbf{else}$
\STATE 10: \hspace{0.55cm}\qquad\qquad $\mathbf{M}\hspace{0.1cm}[i,j] = z'$; $\mathbf{M_d}\hspace{0.1cm}[i,j] = z_{d}'$
\STATE 11: \hspace{0.55cm}\qquad $\mathbf{\hat{I}}\hspace{0.1cm}[i,j]$ =  $\mathbf{I_1}\hspace{0.1cm}[i,j]$ * $\mathbf{M}\hspace{0.1cm}[i,j]$ + $\mathbf{I_2}\hspace{0.1cm}[i,j]$ * $\mathbf{M_d}\hspace{0.1cm}[i,j]$ \\ \qquad\quad\qquad// Step 3: The weight matrix and the image matrix do Hadamard product to generate a new image matrix $\mathbf{\hat{I}}$
\STATE 12: $\mathbf{end}$
\STATE 13: $\mathbf{return}$ $\mathbf{\hat{I}}$
\end{algorithmic}
\label{alg1}
\end{algorithm}

\begin{figure}[htp]
    \centering
    \includegraphics[width=0.49\textwidth]{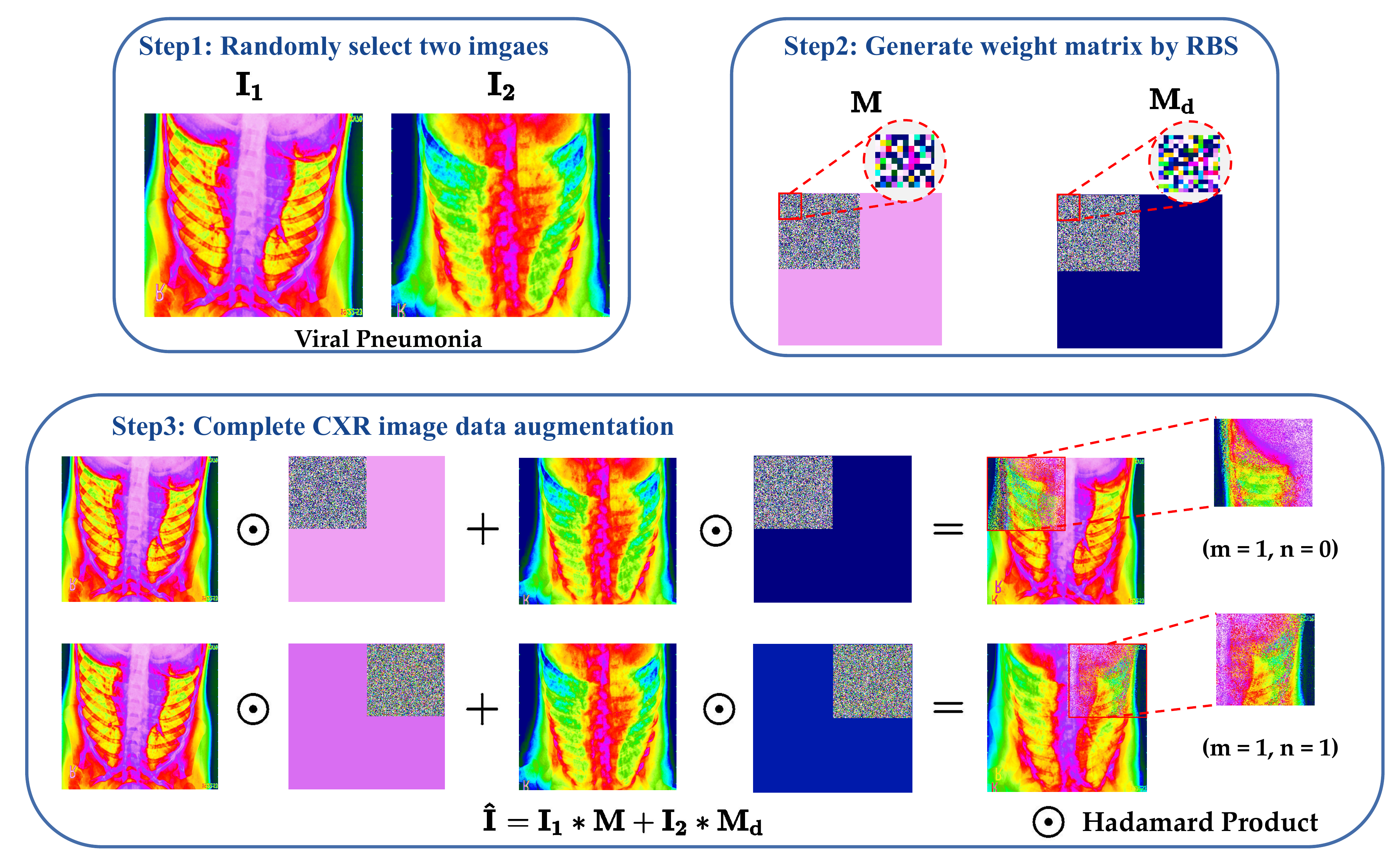}
    \caption{Schematic diagram of the FFA algorithm. Step 1 presents two images randomly selected from the CXR image dataset with viral pneumonia, Step 2 shows the weight matrix generated by RBS when the patch size is 2, and Step 3 shows the two CXR images generated under different position combinations ($m$=1, $n$=0, 1).}
    \label{FFA}
\end{figure}

\subsection{CXR Image Feature Extractor (CIFE)}
\label{CIFE}
The architecture of CIFE is shown in Fig. 3. To extract global and local features from CXR images more comprehensively while maintaining high efficiency. The first four convolutional and pooling layers of pre-trained ResNet101 are selected as the feature extractor. The ResNet is pre-trained based on the ImageNet dataset containing several images of common objects and scenes in life, which cannot be adapted to the medical image classification problem due to the significant differences in the dataset. Therefore, we consider using only its shallow layers (i.e., the first four layers mentioned) with powerful feature extraction capabilities for CXR images and discard the deep layers that are usually classifiers. This means that we can maximize the role of pre-trained ResNet in medical image classification tasks. Given the input CXR image is $\mathbf{I} \in \mathbb{R}^{h\times w \times c}$, the feature maps extracted by CIFE is $\mathbf{I_F} \in \mathbb{R}^{h'\times w' \times c'}$. These feature maps are rich in deep lung features that were encoded as feature embedding $\mathbf{E} \in \mathbb{R}^{C'\times (h' \times w')}$ for input to the subsequent Transformer, which can better understand the relationship between lung features and disease labels.

\begin{figure*}[hpb]
    \centering
    \includegraphics[width=0.75\textwidth]{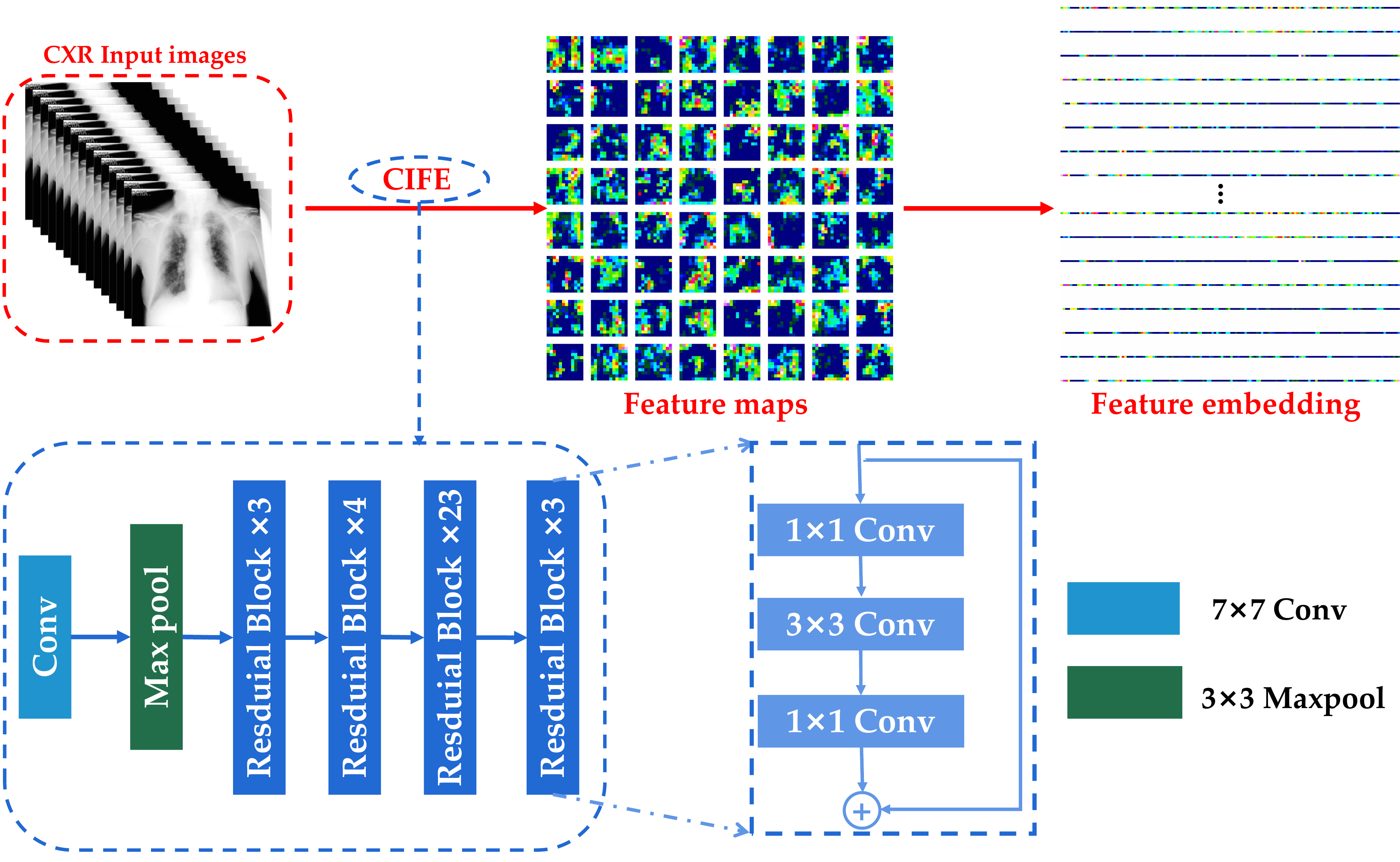}
    \caption{Schematic diagram of the CXR images processing flow and architecture of CIFE. The CXR image features extracted by CIFE are converted into feature embeddings and input to the subsequent Transformer architecture, where "$\times$3", "$\times$4," and "$\times$23," respectively, means that modules were repeated 3, 4, and 23 times.}
    \label{F3}
\end{figure*}

\subsection{Multilevel Mutli-Head Self-Attention Transformer (MMSA-Transformer)}
\label{MT}
The architectuer of the MMSA Transformer is shown in Fig. 4. It mainly consists of a Multilevel Multi-Head Self-Attention (MMSA) used to emphasize the importance of information while ignoring invalid information.
Moreover, a feed-forward network (FFN) is applied to understand further the relationship among task-related feature information by receiving the reconstructed information from the MMSA. FFN then outputs the corresponding probabilities of each label simultaneously rather than in isolation, and the process can be formulated as
\begin{equation}
\mathbf{P}=FFN(\mathbf{X})=Sigmoid((\mathbf{W}^1\mathbf{W}^2\mathbf{X})+b),
\end{equation}
\begin{equation}
Sigmoid(x)=\frac{1}{1+e^{-x}}.
\end{equation}
where the $\mathbf{X}$ is the feature information, the $\mathbf{W}^1$ and $\mathbf{W}^2$ denotes the weight matrix of the first the second layer of FFN, respectively, the $Sigmoid$ is the activation function.

\begin{figure*}[htp]
    \centering
    \includegraphics[width=0.75\textwidth]{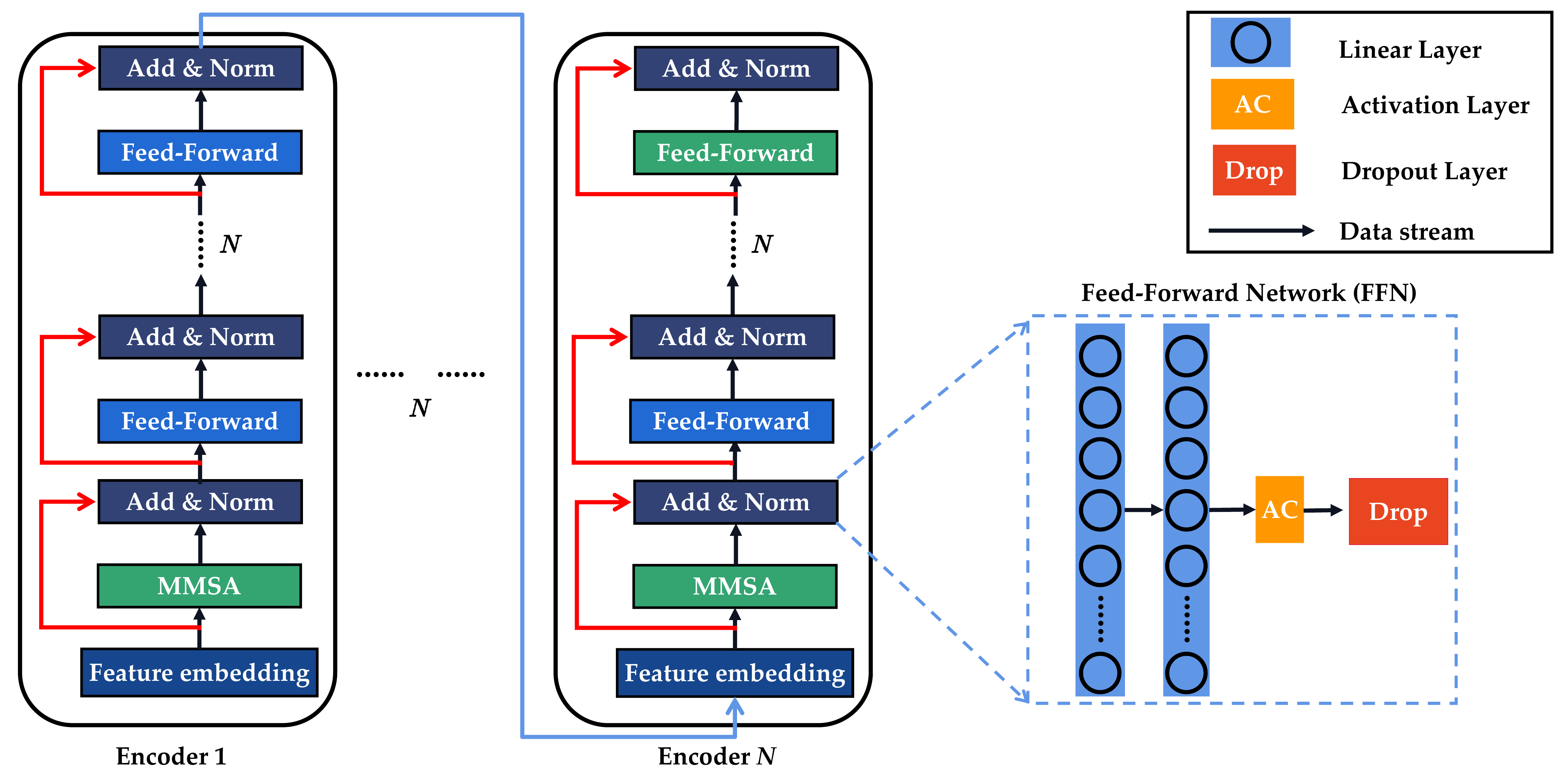}
    \caption{Architecture of MMSA Transformer, where Add \& Norm represent point-wise addition and normalization, MMSA is the proposed Multilevel Multi-Head Attention, and the $N$ denotes the number of Encoder and the repeated times of 'Feed-Forward-Add \& Norm' sequence.}
    \label{F5}
\end{figure*}

Traditionally, the Transformer-based models rely heavily on the Multi-Head Self-Attention (MHSA) to capture the critical information on data, as shown in \textbf{Left} of Fig. 5.
However, the high computational complexity of this mechanism inevitably increases the model's inference time while obtaining excellent performance. To reduce the computational complexity of the attention mechanism, so that satisfy the high-speed recognition requirement while keeping outstanding accuracy, we design MMSA, an highly efficient and practical self-attention mechanism, as shown in \textbf{Right} of Fig. 5. Next, we first briefly introduce the conventional MHSA, while our proposed MMSA is presented and analyzed afterward.

\begin{figure*}[htp]
    \centering
    \includegraphics[width=0.75\textwidth]{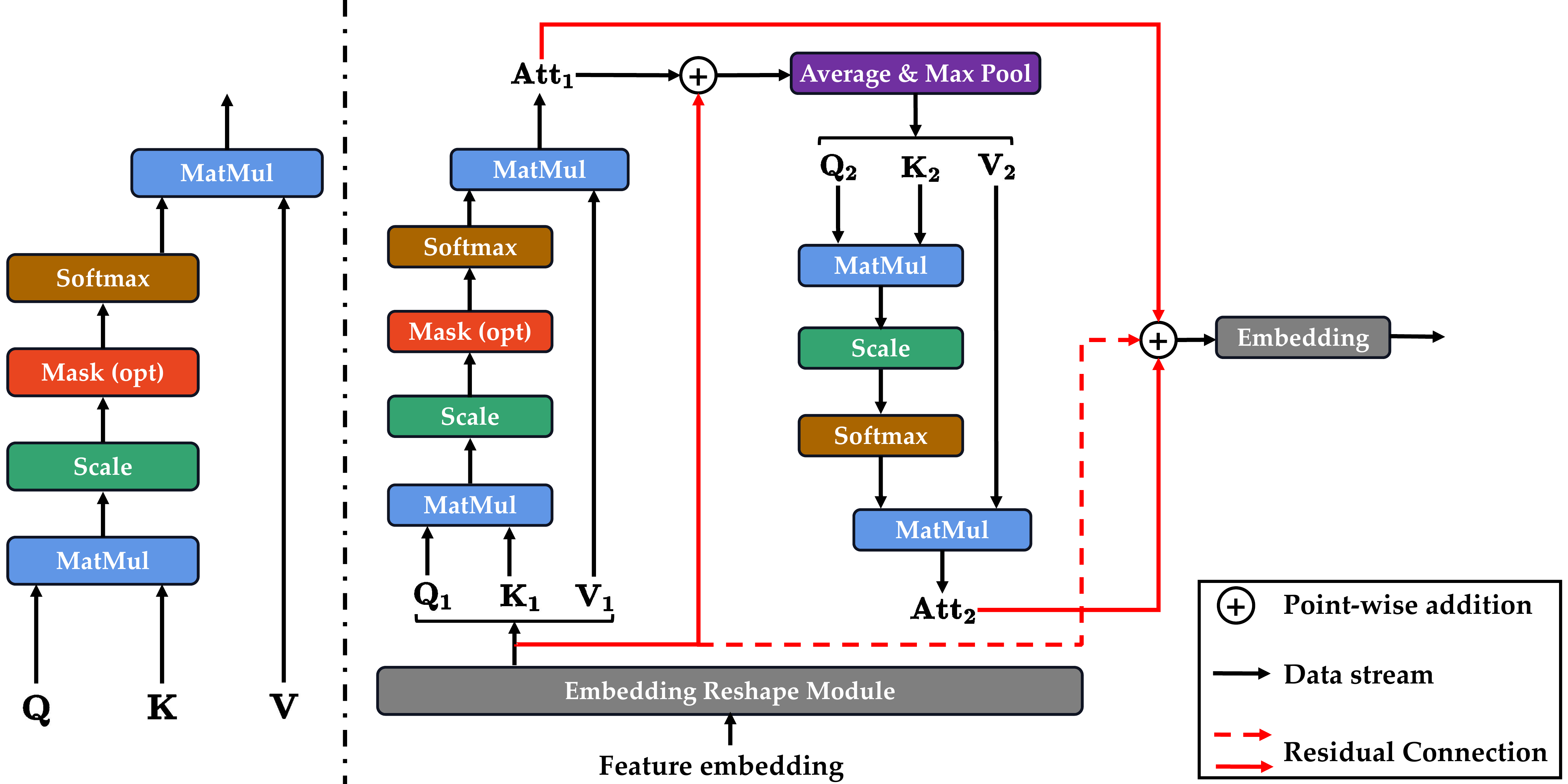}
    \caption{Architecture of Multi-Head Self-Attention. \textbf{(Left)} Conventional Multi-Head Self-Attention (MHSA); \textbf{(Right)} The proposed Mutlilevel Multi-Head Self-Attention (MMSA), in which the embedding reshape module revice the feature embedding from CIFE.}
    \label{F5}
\end{figure*}

\subsubsection{Multi-Head Self-Attention (MHSA)}
Given the feature map $\mathbf{I_F} \in \mathbb{R}^{c \times h \times w}$, where $c$, $h$, and $w$ denotes the channels, height, and width, respectively. The $\mathbf{I_F}$ was reshaped before fed into the attention module to define and obtain the query $\mathbf{Q}$, key $\mathbf{K}$, and value $\mathbf{V}$, these can be formulated as follow: 
\begin{equation}
\mathbf{I_F} \in \mathbb{R}^{c \times h \times w} \xrightarrow{reshape} \mathbf{I_F} \in \mathbb{R}^{c \times (h \times w)},
\end{equation}
\begin{equation}
\mathbf{Q}=W^q\mathbf{I_F}, \mathbf{K}=W^k\mathbf{I_F}, \mathbf{V}=W^v\mathbf{I_F}, 
\end{equation}
where the $W^q$, $W^k$, and $W^v$ denotes the trainable weight matrices for $\mathbf{Q}$, $\mathbf{K}$, and $\mathbf{V}$, respectively. The attention matrix can be obtained from $\mathbf{Q}$ and $\mathbf{K}$, that can be presented as
\begin{equation}
\mathbf{Att}=Softmax(\mathbf{QK}^T/\sqrt{d}),
\end{equation}
where the $\sqrt{d}$ is an approximate normalization constant. Then the output obtain from $\mathbf{Q}$, $\mathbf{K}$, and $\mathbf{V}$ can be expressed as:
\begin{equation}
\mathbf{A}=\mathbf{Att}\mathbf{V}.
\end{equation}
The matrix product $\mathbf{QK}^T$ first computes the similarity between each pair of tokens. Then, based on the similarity, each new token is derived by the combination of all tokens. Finally, the attention's output can be computed by a residual connection, which can be formulated as
\begin{equation}
\mathbf{I_F} \in \mathbb{R}^{c \times (h \times w)} \xrightarrow{reshape} \mathbf{I_F} \in \mathbb{R}^{c \times h \times w},
\end{equation}
\begin{equation}
\mathbf{I_o} = \mathbf{A}\mathbf{W}^p,
\end{equation}
in which the $\mathbf{W}^p$ is a trainable matrix that shape is $\mathbf{W}^p \in \mathbb{R}^{c \times c}$ for feature projection. Given that this attention module contains $N$ heads internally, i.e., $N$-heads Self-Attention, its final attention output is then summed over the channel dimensions according to the results of Eq.(9).

The computation complexity of Multi-Head Self-Attention can be inferred according the Eq.(4) - (9):
\begin{equation}
\Omega(\text{MHSA})=2ch^2w^2+3hwc^2.
\label{mhsa}
\end{equation}
The space complexity includes the term $O(h^2 w^2)$, which means that the memory consumption could become very large when the input is high-resolution ($h \times w$). This limits the application of the MHSA module to CXR image recognition tasks that place demands on fast training and inference. Based on this problem, we propose MMSA, a multilevel multi-head self-attention mechanism with low computational complexity. It reduces the complexity while obtaining an excellent information understanding capability under different levels without compromising performance.

\subsubsection{Multilevel Multi-Head Self-Attention (MMSA)}
\label{MMSA}
The architecture of the proposed MMSA is shown in the \textbf{Right} of Fig. 5. Compared with the conventional MHSA, MMSA does not directly calculate the attention of the whole input image but pre-divides the input into different levels, which means that only a limited number of tokens need to be calculated in the attention calculation of each level. This form of attention calculating can encourage the model to comprehensively focus on more fine-grind feature regions and capture the relation of global-local information instead of paying all attention to overall input. Given the input feature map $\mathbf{I_F} \in \mathbb{R}^{c \times h \times w}$, it will be divided into several small grid according to its original resolution before calculating the attention. The small grid can be considered as $\mathbf{G} \in \mathbb{R}^{g \times g}$, and the input feature map be reshaped according to the small grid as:
\begin{equation}
\begin{aligned}
\mathbf{I_F} \in \mathbb{R}^{c \times h \times w} \rightarrow \mathbf{I_{F1}} \in \mathbb{R}^{c \times (\frac{h}{g} \times g) \times (\frac{w}{g} \times g)} \\
\rightarrow \mathbf{I_{F1}} \in \mathbb{R}^{c \times (\frac{h}{g} \times \frac{w}{g}) \times (g \times g) }.
\end{aligned}
\end{equation}
The query, key, and value of this level can be obtained by
\begin{equation}
\mathbf{Q_1}=\mathbf{I_{F1}}\mathbf{W}^{q}_{1}, \mathbf{K_1}=\mathbf{I_{F1}}\mathbf{W}^{k}_{1}, \mathbf{V_1}=\mathbf{I_{F1}}\mathbf{W}^{v}_{1},
\end{equation}
where $\mathbf{W}^{q}_{1}$, $\mathbf{W}^{k}_{1}$, and $\mathbf{W}^{v}_{1}$ are trainable weight matrices that is $\mathbf{W}^{q}_{1} \in \mathbb{R}^{c \times c}$, $\mathbf{W}^{k}_{1} \in \mathbb{R}^{c \times c}$, and $\mathbf{W}^{v}_{1} \in \mathbb{R}^{c \times c}$, respectively. Then, the local attention of this level can be generated by
\begin{equation}
    \mathbf{Att_1}=Softmax(\mathbf{Q_{1} K_{1}}^T/\sqrt{d})\mathbf{V_1},
\end{equation}
$\mathbf{Att_1}$ is reshaped to the shape of the input image $\mathbf{I_F}$ before computing the next level of local attention, and a residual connection applied between this local attention and the input image, which can be formulated as
\begin{equation}
\begin{aligned}
\mathbf{Att_1} \in \mathbb{R}^{c \times (\frac{h}{g} \times \frac{w}{g}) \times (g \times g)} &\rightarrow \mathbf{Att_1} \in \mathbb{R}^{c \times (\frac{h}{g} \times g) \times (\frac{w}{g} \times g)} \\
&\rightarrow \mathbf{Att_1} \in \mathbb{R}^{c \times h \times w },
\end{aligned}
\end{equation}
\begin{equation}
\mathbf{Att_1} = \mathbf{Att_1} + \mathbf{I_F}.
\end{equation}
Compared with the conventional Multi-Head Self-Attention that operates directly on $h$ or $w$, $\mathbf{Att_1}$ directly performs the attention operation on the small grid $g \times g$, which reduces its space complexity significantly.

After entering the next computational level, $\mathbf{Att_1}$ is downsampled to $g'$ by pooling. Consistent with the calculation of the first level, each small grid $\mathbf{G'} \in \mathbb{R}^{g' \times g'}$ will be regarded as a token in the calculation, and the process can be expressed as 
\begin{equation}
\mathbf{Att_1'} = \frac{1}{2}[\alpha \times MaxPool_{g'}(\mathbf{Att_1}) + \beta \times AvePool_{g'}(\mathbf{Att_1})],
\end{equation}
Here, the $MaxPool_{g'}(x)$ and $AvePool_{g'}(x)$ denotes to downsample the input feature map by $g'$ times using max pooling and average pooling with the filter size and stride of $g'$, respectively, and the $\alpha$ and $\beta$ are control coefficients, they are both positive greater than 0 and add to 1. From this, it can be obtained that the $\mathbf{Att_1'}$ after downsampling is $\mathbf{Att_1'} \in \mathbf{R}^{c \times \frac{h}{g'} \times \frac{w}{g'}}$. Furthermore, the token size $(\frac{h}{g'} \times \frac{w}{g'})$ can be obtained form the reshaped $\mathbf{Att_1'}$:
\begin{equation}
\mathbf{Att_1'} \in \mathbf{R}^{c \times \frac{h}{g'} \times \frac{w}{g'}} \rightarrow \mathbf{Att_1'} \in \mathbf{R}^{c \times (\frac{h}{g'} \times \frac{w}{g'})}.
\end{equation}
The query, key, and value of this level can be computed by $\mathbf{Att_1'}$ as
\begin{equation}
\mathbf{Q_2}=\mathbf{Att_1'}\mathbf{W}^{q}_{2}, \mathbf{K_2}=\mathbf{Att_1'}\mathbf{W}^{k}_{2}, \mathbf{V_2}=\mathbf{Att_1'}\mathbf{W}^{v}_{2},
\end{equation}
where $\mathbf{W}^{q}_{2}$, $\mathbf{W}^{k}_{2}$, and $\mathbf{W}^{v}_{2}$ are trainable weight matrices that is $\mathbf{W}^{q}_{2} \in \mathbb{R}^{c \times c}$, $\mathbf{W}^{k}_{2} \in \mathbb{R}^{c \times c}$, and $\mathbf{W}^{v}_{2} \in \mathbb{R}^{c \times c}$, respectively. Next, the local attention $\mathbf{Att_2}$  can be obtained by
\begin{equation}
    \mathbf{Att_2}=Softmax(\mathbf{Q_{2} K_{2}}^T/\sqrt{d})\mathbf{V_{2}},
\end{equation}

The final output of MMSA can be computed by the reshaping operation and a residual connection:
\begin{equation}
\mathbf{Att_2} \in \mathbb{R}^{c \times (h \times w)} \rightarrow \mathbf{Att_2} \in \mathbb{R}^{c \times h \times w},
\end{equation}
\begin{equation}
\mathrm{MMSA}(\mathbf{I_F}) = (\mathbf{Att_1} + \mathbf{Att_2})\mathbf{W}^m \mathbf{W}^n + \mathbf{I_F},
\end{equation}
where $\mathbf{W}^{m}$, $\mathbf{W}^{n}$ are trainable weight matrices. The proposed MMSA has strong local and global feature modeling capabilities by fusing different levels of attention outputs. Next, the computation complexity of MMSA will be analyzed.
The computational complexity of MMSA can be computed by
\begin{equation}
\Omega(\text{MMSA})=chw(2g'^2 + 4c) + \frac{2chw}{g'^2}(c+hw),
\label{mmsa}
\end{equation}
which means that the computation complexity from $O(h^2 w^2)$ reduce to $O(hwg^2 + \frac{h^2 w^2}{g'^2})$ .
\subsection{Computational Complexity Analysis}
Eq.(\ref{mhsa}) and Eq.(\ref{mmsa}) show the computational complexity of the conventional MHSA and our MMSA, respectively. Suppose a CXR image is $360\times360\times3$, and $g'$ in MMSA is 2, the computational parameters of conventional MHSA and our MMSA are:
\begin{equation}
\begin{aligned}
  \text{P}_{\text{MHSA}} = 2*3*360^4+3*360^2*3^2 = 100,780,459,200,\\
  \text{P}^{g'=2}_{\text{MMSA}} = 3*360^2*(2^3+4*3) + \frac{2*3*360^2}{2^2}*(3+360^2) \\ = 25,202,599,200,\\
    \text{P}^{g'=5}_{\text{MMSA}} = 3*360^2*(5^3+4*3) + \frac{2*3*360^2}{5^2}*(3+360^2) \\ = 4,084,437,312,\\
    \text{P}^{g'=10}_{\text{MMSA}} = 3*360^2*(10^3+4*3) + \frac{2*3*360^2}{10^2}*(3+360^2) \\ = 1,401,258,528. 
\end{aligned}
\end{equation}
This menas that the proposed MMSA pays only a quarter of the computational cost of the conventional MHSA when $g'=2$. In addition, the $\text{P}_{\text{MMSA}}$ is far smaller than the $\text{P}_{\text{MHSA}}$. The relationship between the conventional MHSA and our MMSA with different parameters $g'$ can be formulated as:
\begin{equation}
    \text{P}_{\text{MHSA}} \approx 4\times\text{P}^{g'=2}_{\text{MMSA}} \approx 25\times\text{P}^{g'=5}_{\text{MMSA}} \approx 72\times\text{P}^{g'=10}_{\text{MMSA}}.
\label{relat}
\end{equation}
According to Eq.(\ref{relat}), when the $g'=2$, the conventional MHSA has at least four times more parameter operations than the MHSA. In addition, when $g'=5, 10$, this number reaches a staggering 25 and 72 times.
\section{Experiments}
\label{EXP}
This experiments section shows the details of dataset, the experimental setup, evaluation metrics, and analyzes and discusses the experimental results, and model interpretability exploration.
\subsection{Dataset}
\begin{figure}[tbh]
    \centering
    \includegraphics[width=0.7\textwidth]{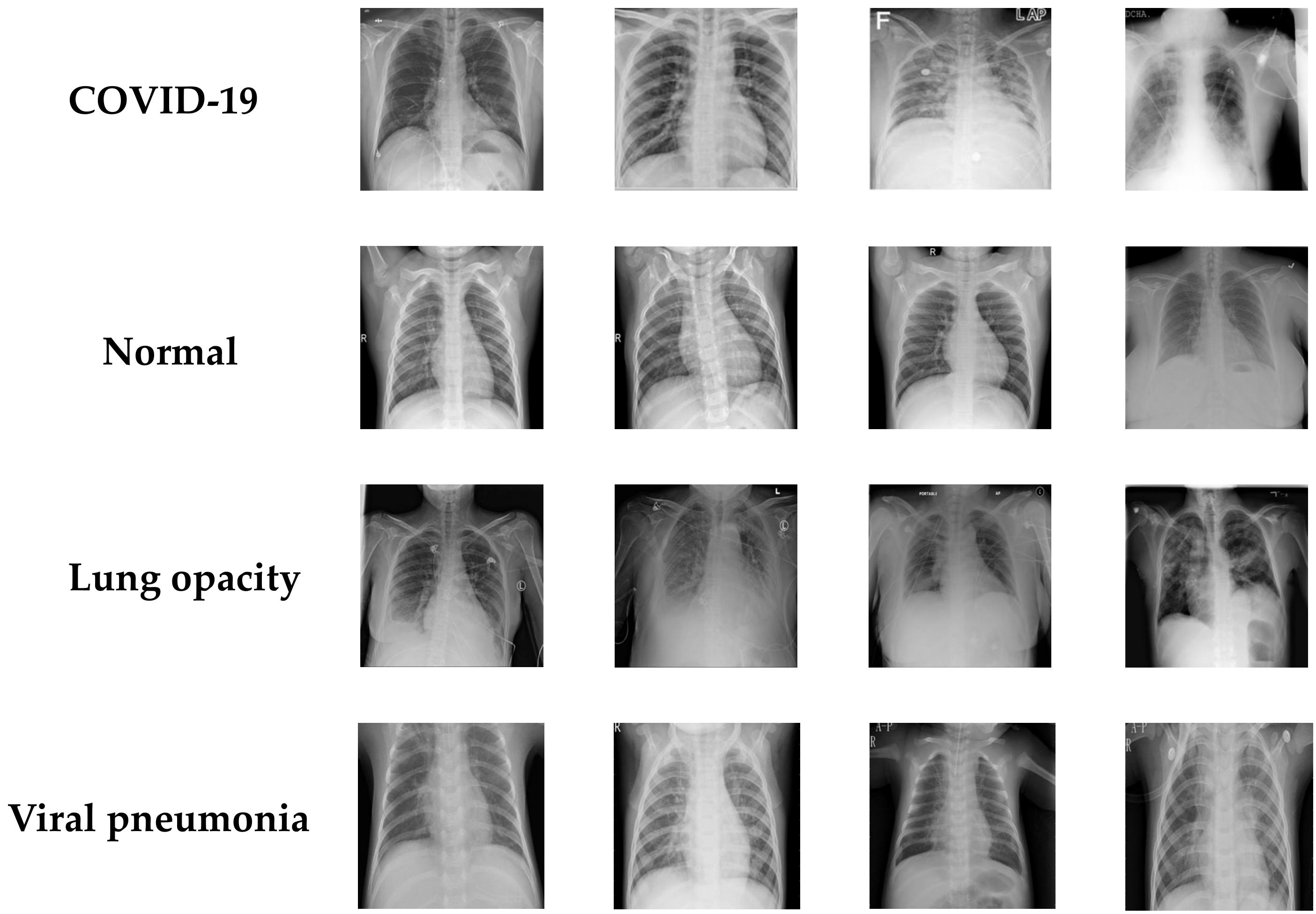}
    \caption{Examples of CXR image datasets, from top to bottom, are COVID-19 positive, Normal, Lung Opacity, and Viral pneumonia.}
    \label{DatasetExample}
\end{figure}

The pneumonia dataset used in our work consists of CXR images corresponding to different symptoms: Normal, COVID-19, Lung Opacity, and Viral pneumonia~\cite{chowdhury2020can, rahman2021exploring}. The dataset has 21,175 CXR images with $299 \times 299$ resolution. These CXR images were collected from five different medical institutions such as Banco digital de Imagen Médica de la Comunidad Valenciana (BIMCV), the German Medical School (GMS), the Società Italiana di Radiologia Medica e Interventistica (SIRM), Radiological Society of North America (RSNA), and Guangzhou Women's and Children's Medical Center (GWCMC). The detailed information of the dataset is presented in Table~\ref{T1}, and some exampled images are shown in Fig.~\ref{DatasetExample}.

\begin{table}[htbp]
  \centering
  \caption{Detailed information of the pneumonia recognition dataset, which source, amount, resolution, and total represent the data source, number of data provided by data sources, image resolution, and the total number of images, respectively.}
    \begin{tabular}{ccccc}
    \toprule
    Disease & Data Source & Amount & Resolution & Total \\
    \midrule
    \midrule
    \multirow{3}[1]{*}{COVID-19} & BIMCV & 2473 & \multirow{3}[1]{*}{299 × 299} & \multirow{3}[1]{*}{3615} \\
       & GMS & 183 &  \\
       & SIRM & 959 &  \\
    \midrule
    \multirow{2}[0]{*}{Normal} & RSNA & 8851 & \multirow{2}[0]{*}{299 × 299} & \multirow{2}[0]{*}{10192} \\
       & GWCMC & 1341 &  \\
    \midrule
    Lung Opacity & RSNA & 6012 & 299 × 299 & 6023 \\
    \midrule
    Viral pneumonia & GWCMC & 1345 & 299 × 299 & 1345 \\
    \bottomrule
    \end{tabular}%
  \label{T1}%
\end{table}%

\subsection{Experimental Setup}
All models and algorithms in this work are built on the Pytorch. The first four layers of ResNet101, pre-trained by ImageNet, are considered the CIFE backbone network for accelerating the convergence of the overall model. The MMSA Transformer is trained from scratch by receiving feature embeddings from CIFE to enter the sage. In the pre-training FFA module, the patch size ($p$) is set to 2, $\alpha$ to 0.4, and the position parameters $m$, $n$ are set to the permutations obtained from $(m=0, 1; n=0,1)$, such as $m=1,n=1$ or $m=0,n=1$, etc. During training, $N$ in the MMSA Transformer structure is set to $N=4$ and $g$ and $g'$ in the MMSA are set to [2, 2], and $\alpha$ and $\beta$ in Eq. (18) are set to 0.3 and 0.7, respectively. The first and second momentum is 0.9, and 0.999 of Adam\cite{kingma2014adam} were considered as optimizers of the network to minimize the loss function with a learning rate of $10^{-5}$. 
During the training process, the CXR image dataset is randomly disrupted and then divided into training, test, and validation sets in the ratio of 8 (16,940) :1 (2,117) :1 (2,117), without any overlap. Note that all results generated by the above model are based on twenty iterations of model training, and the early stopping strategy is adopted.
The learning rate Warmup mechanism \cite{he2016deep} and Cosine Annealing strategy \cite{loshchilov2016sgdr} is employed to avoid model oscillation and accelerate convergence, which can be expressed as
\begin{equation}
\eta_t=\eta_{\min }^i+\frac{1}{2}\left(\eta_{\max }^i-\eta_{\min }^i\right)\left(1+\cos \left(\frac{T_{c u r}}{T_i} \pi\right)\right),
\end{equation}
where $\eta^i_{max}$ and $\eta^i_{min}$ is the max and min learning rate that define the range of learning rate, $i$ denotes the index of iteration times, and $T_{cur}$ and $T_{i}$ is the current iteration times and total iteration times. The loss function is BCE Loss, which can be formulated as
\begin{equation}
L(o, t)=-1 / n \sum_i(t[i] * \log (o[i])+(1-t[i]) * \log (1-o[i])),
\end{equation}
Here, the $o$ and $t$ are the output and actual target, respectively. In addition, in each FFN within the MMSA-Transformer, a Dropout operation with probability 0.2 is applied to avoid overfitting the model. All images are adjusted to 360 × 360 and converted to RGB images before input into the model, and each mini-batch contains 16 randomly scrambled images.

All the algorithms and models run on the Tesla V100 GPU (16 GB), Intel Xeon W-2245 (3.90 GHz).

\subsection{Evaluation Metrics}
The recall ($R$) and precision ($P$) of each disease label were selected as evaluation metrics. The results were first classified into the true positive (TP), true negative (TN), false positive (FP), and false negative (FN) according to the classification, as shown in Table II. The $R$ and $P$ can be formulated as

\begin{equation}
\begin{aligned}
R &= \frac{TP}{TP+FN}, \\
P &= \frac{TP}{TP+FP},
\end{aligned}
\end{equation}
Then, the average class precision (CP) and the average class recall (CR), which denotes the average of each class precision and recall, are calculated by
\begin{equation}
\begin{aligned}
C R &=\frac{\sum_{n=1}^N \sum_{n=1}^K R}{NK}, \\
CP &=\frac{\sum_{n=1}^N \sum_{n=1}^K P}{NK},
\end{aligned}
\end{equation}
where $N$ and $K$ is the scale of the test dataset and the number of disease labels, respectively. Similarity, the overall precision (OP) and overall recall (OR) that means the actual prediction of all images in all disease labels, respectively, can be formulated as
\begin{equation}
\begin{aligned}
O R &=\frac{\sum_{n=1}^N \sum_{n=1}^K f(p_{n,i},\hat{p}_{n,i})}{\sum_{n=1}^N \sum_{n=1}^K p_{n,i}}, \\
CP &=\frac{\sum_{n=1}^N \sum_{n=1}^K f(p_{n,i},\hat{p}_{n,i})}{NK},
\end{aligned}
\end{equation}
where $N$ and $K$ is the scale of the test dataset and the number of disease labels, respectively, the $p_{n, i}$ and $\hat{p}_{n,i}$ is the actual label and prediction of the $n^{th}$ image on the $i^{th}$ disease, respectively. The function $f(\cdot)$ is computed by
\begin{equation}
f(p, \hat{p})=\left\{\begin{array}{lc}
1, & p=\hat{p} \\
0, & \text { otherwise }
\end{array}\right.
\end{equation}
The CF1 and OF1 can be computed according to Eq.(23) - (25), which can be formulated as
\begin{equation}
\begin{aligned}
CF1 &= \frac{2\times CR\times CP}{CR+CP}, \\
OF1 &= \frac{2\times OR\times OP}{OR+OP}.
\end{aligned}
\end{equation}

\subsection{Baseline}
The baseline in our work is divided into two distinct parts: Nonlinear Classifiers and Deep Neural Networks.
agraph{Nonlinear Classifiers} Three representative nonlinear classifiers, such as Decision Tree (DT), Support Vector Machine (SVM), and Multilayer Perceptron (MLP), are selected as baselines of the nonlinear classifiers part. The Principal Component Analysis (PCA) reduces input CXR images' dimensions before these classifiers operate. Moreover, the kernel function of the SVM is RBF, and all hidden layers' unit is 512 in MLP.
agraph{Deep Neural Networks} Currently popular CNN-based models such as MobileNet~\cite{howard2017mobilenets}, AlexNet~\cite{krizhevsky2012imagenet}, VGGNet~\cite{simonyan2014very}, ResNet~\cite{he2016deep}, ResNeXt~\cite{xie2017aggregated}, Wide ResNet~\cite{zagoruyko2016wide}, Incepition V3~\cite{szegedy2017inception}, ConvNeXt~\cite{liu2022convnet}, and RegNe~\cite{radosavovic2020designing}, pre-trained by ImageNet~\cite{deng2009imagenet}, were selected as baselines for comparison. Specially, there is VGGNet16, ResNet101, ResNeXt101-32×8d, Wide ResNet101, Inception V3, ConvNeXt-base, and RegNetY-400MF. In addition, CNN-RNN-based models (e.g., CNN-LSTM and CNN-ConvLSTM) are considered baselines in which the CNN is ResNet152 pre-trained by ImageNet, and the LTSM/ConvLSTM is a four-layers architecture. Note that the final classification layer of these CNN-based backbones was changed to four when those acts separately to fit the recognition task in our work. Moreover, the CNN-Transformer with the conventional MHSA is considered the baseline to demonstrate the advantage of MMSA, the number of attention heads and Transformer layers is consistent with the proposed CMT.

\subsection{Experimental Results}
Three machine learning-based nonlinear classifiers and several deep neural network-based classifiers were utilized to validate the superiority of our proposed CMT. The experimental results section is illustrated in three parts: Nonlinear Classifiers, Deep Neural Network-based Classifiers, and Training and Inference Time.
agraph{Comparison with Nonlinear Classifiers} The performance comparison of the proposed CMT with nonlinear classifiers is presented in Table~\ref{nonlinear}. The proposed CMT outperforms a series of nonlinear classifiers that act separately to achieve recognition. Moreover, to validate the superiority of CMT in a more fair way, the pre-trained ResNet101 is adopted as the dimensional down-sampler spliced with these nonlinear classifiers as the extra baseline for comparison. Clearly, the proposed CMT still has better recognition performance than these extra baselines (average higher $8.1\%$, $6.0\%$, $4.4\%$ than ResNet \& DT, SVM, MLP, respectively), although the pre-trained ResNet significantly improves original nonlinear classifiers (average $20.8\%$, $16.1\%$, $17.2\%$ in DT, SVM, MLP, respectively). This means our MMSA Transformer keeps excellent performance and is far beyond the conventional classifiers. The advantages of our proposed CMT over conventional nonlinear classifiers are verified.
\begin{table}[tbh]
  \centering
  \caption{Performance comparison of the proposed CMT with nonlinear classifiers, the ResNet \& (DT/SVM/MLP) denotes which the pre-trained ResNet101 as the dimensional down-sampler and DT/SVM/MLP as the classifier.}
  \resizebox{0.9\textwidth}{!}{
    \begin{tabular}{ccccc}
    \toprule
    Model & COVID-19 & Normal & Lung Optiacy & Viral Pneumonia \\
    \midrule
    DT & 0.7046 & 0.7767 & 0.7447 & 0.7282 \\
    SVM & 0.7522 & 0.7606 & 0.8284 & 0.7966 \\
    MLP & 0.7866 & 0.8026 & 0.8193 & 0.7473 \\
    \midrule
    ResNet \& DT & 0.9361 & 0.8444 & 0.8767 & 0.9141 \\
    ResNet \& SVM & 0.9524 & 0.8458 & 0.9395 & 0.9047 \\
    ResNet \& MLP & 0.9453 & 0.9023 & 0.9224 & 0.9285 \\
    CMT (Ours) & \textbf{0.9972} & \textbf{0.9628} & \textbf{0.9714} & \textbf{0.9303} \\
    \bottomrule
    \end{tabular}}
  \label{nonlinear}%
\end{table}%
agraph{Comparison with Deep Neural Network-based Classifiers}
The experiment results of performance comparison between CMT and deep neural network-based baseline are shown in \textbf{Table~\ref{deepneural}}. We can observe that \textbf{i)} The overall recognition accuracy and generalization performance of the proposed CMT outperforms all baselines; \textbf{ii)} the F1 scores (CF1 and OF1) of the CMT are $6.4\%$ and $7.9\%$ higher than the CNN-Transformer, respectively, which means that the proposed MMSA mechanism has a significant advantage over conventional MHSA. \textbf{iii)} our proposed CMT, with ResNet101 as the backbone on default, has better recognition performance than its variants (e.g., the backbone is replaced with Inception V3/VGGNet/ResNeXt/Wide Resnet/ConvNeXt/RegNet), in terms of recognition accuracy and evaluation metrics' scores. The proposed CMT's average recognition accuracy is $2.4\%$, $3.1\%$, $1.3\%$, $2.1\%$, $3.3\%$, $2.4\%$ higher than the ones that backbone is Inception V3/VGGNet/ResNeXt/Wide Resnet/ConvNeXt/RegNet. The average F1 score is $5.1\%$, $4.1\%$, $7.1\%$, $5.8\%$, $5.4\%$, $4.0\%$ higher than the CMT with the Inception V3/VGGNet/ResNeXt/Wide Resnet/ConvNeXt/RegNet backbone. In addition, the performance of CMT's variants with different backbones is far beyond these pre-trained backbone act independently, which means the proposed MMSA Transformer within CMT can enhance CNN-based models. The superiority of MMSA and CMT, the advantage of our MMSA over MHSA, and the necessity of ResNet101 as the default backbone is validated.
agraph{Training and Inference Time} In addition, to verify the superiority of the proposed MMSA mechanism in terms of computational complexity, we experiment with comparing training and inference timing between the CNN-Transformer with conventional MHSA, Windows Multi-Head Self Attention (W-MSA)~\cite{liu2021swin}, Pale-Shaped Attention (PS)~\cite{wu2022pale}, and Global Self-Attention (GSA)~\cite{shen2020global} and CMT (with proposed MMSA). The result is shown in \textbf{Table~\ref{computation}}. When keeping the same model parameters, the proposed CMT consumes significantly less time for computing than the CNN-Transformer, is $38.2\%$/$28.1\%$/$18.9\%$/$6.8\%$ and $29.4\%$/$19.4\%$/$11.1\%$/$4.2\%$ faster than the CNN-Transformer with MHSA/W-MSA/PS/GSA in terms of training and inference, respectively. This means that the proposed MMSA mechanism reduces the computational complexity while maintaining an excellent recognition performance.

\begin{table*}[htbp]
  \centering
  \caption{Experimental result on the CXR image dataset, the per-class results (COVID-19/Normal/Lung Optiacy/Viral pneumonia) is the recognition accuracy, and the \textbf{bold} represents the optimal value in each evaluation metric.}
  \resizebox{\textwidth}{!}{
    \begin{tabular}{ccccccccccc}
    \toprule
    Model & COVID-19 & Normal & Lung Optiacy & Viral pneumonia & CP & CR & CF1 & OP & OR & OF1 \\
    \midrule
    \midrule
    MoblieNet~\cite{howard2017mobilenets} & 0.8934 & 0.9426 & 0.8833 & 0.8412 & 89.0 & 83.2 & 86.0 & 90.4 & 74.1 & 81.5 \\
    AlexNet~\cite{krizhevsky2012imagenet} & 0.7727 & 0.8802 & \textbf{0.9729} & 0.7560 & 87.5 & 84.4 & 85.9 & 87.2 & 81.7 & 84.4 \\
    ResNet~\cite{he2016deep} & 0.9211 & 0.8366 & 0.9148 & 0.9018 & 82.1 & 89.4 & 85.6 & 87.2 & 82.3 & 84.6 \\
    VGGNet~\cite{simonyan2014very} & 0.9466 & 0.8558 & 0.8537 & 0.7642 & 88.0 & 82.7 & 85.3 & 89.9 & 73.2 & 80.7 \\
    ResNeXt~\cite{xie2017aggregated} & 0.9418 & 0.8885 & 0.8809 & 0.8569 & 92.5& 84.6 & 88.4 & 81.2 & 85.9 & 83.5\\
    Wide ResNet~\cite{zagoruyko2016wide} & 0.9190 & 0.9037 & 0.8406 & 0.8542 & 92.9 & 85.7 & 89.2 & 82.7 & 82.5 & 82.6 \\
    Incepition V3~\cite{szegedy2017inception} & 0.9552 & 0.8912 & 0.9018 & 0.9145 & 86.2 & 84.1 & 85.1 & 83.7 & 83.3 & 83.5\\
    ConvNeXt~\cite{liu2022convnet} & 0.9677 & 0.9090 & 0.8563 & 0.8671 & 90.4& 81.8 & 85.9 & 90.0 & 81.2 & 85.4 \\
    RegNet~\cite{radosavovic2020designing}  & 0.9721 & 0.9624 & 0.8515 & 0.7904 & 87.7 & 81.1 & 84.2 & 88.4 & 80.8 & 84.4 \\
    CNN-LSTM & 0.8121 & 0.9088 & 0.9038 & 0.7820 & 85.2 & 89.2 & 87.1 & 88.7 & 87.3 & 88.0 \\
    CNN-ConvLSTM & 0.8702 & 0.9058 & 0.9431 & 0.8821 & 90.0 & 88.6 & 89.3 & 90.5 & 85.9 & 88.2 \\
    CNN-Transformer & 0.9806  & 0.8611  & 0.8992  & 0.9308  & 89.1  & 94.3  & 91.6  & 93.8  & 86.6  & 90.0 \\
    \midrule
    CMT (Inception V3) & 0.9841  & 0.8662  & 0.9292  & 0.9289  & 90.8  & 89.9  & 90.3  & 94.2  & 95.7  & 94.9 \\
    CMT (VGGNet) & 0.9624  & 0.9002  & 0.9700  & 0.9123  & 91.7  & 95.9  & 93.8  & 96.6  & 90.0  & 93.2 \\
    CMT (ResNeXt) & 0.9787 & 0.9608 & 0.9203 & 0.9517 & 95.3 & 87.8 & 91.4 & 94.2 & 86.7 & 90.3 \\
    CMT (Wide ResNet) & 0.9625 & 0.9029 & 0.9497 & \textbf{0.9687} & 94.6 & 88.1 & 91.2 & 93.9 & 86.5 & 90.1\\
    CMT (ConvNeXt) & 0.9724& 0.9212 & 0.9421 & 0.9012 & 94.1 & 90.0 & 92.0 & 93.7 & 91.6 & 92.6\\
    CMT (RegNet) & 0.9779 & 0.9620 & 0.9354 & 0.8975 & 95.5 & 91.4 & 93.4 & 95.5 & 92.2 & 93.8\\
    CMT (Ours, ResNet101)& \textbf{0.9972}  & \textbf{0.9628}  & 0.9714  & 0.9303  & \textbf{96.9}  & \textbf{98.2}  & \textbf{97.5}  & \textbf{97.0}  & \textbf{97.3}  & \textbf{97.1} \\
    \bottomrule
    \end{tabular}}
  \label{deepneural}%
\end{table*}%

\begin{table}[tbh]
  \centering
  \caption{Training and inference time of \textbf{one epoch} between the proposed MMSA and the conventional MHSA with the same number of parameters, and the \textbf{bold} represents the optimal value. The device is Tesla V100 GPU (16 GB).}
  \resizebox{.8\textwidth}{!}{
    \begin{tabular}{cccccc}
    \toprule
    Model & Attention & Param\# & Training  & Inference\\
    \midrule
    CNN-Transformer & MHSA & 195.728 M & 738 s & 102 s\\
    CNN-Transformer & W-MSA & 195.728 M & 584 s & 86 s \\
    CNN-Transformer & PS & 195.728 M & 542 s & 80 s \\
    CNN-Transformer & GSA & 195.728 M & 487 s & 75 s \\
    CMT (Ours) & MMSA & 195.728 M & \textbf{456 s} & \textbf{72 s} \\
    \bottomrule
    \end{tabular}}
  \label{computation}%
\end{table}%

\subsection{Ablation Experiments}
Extensive ablation studies are conducted to validate the effectiveness and necessity of the proposed FFA and MMSA. The model's hyperparameters setting is kept the same as the previous experiments in the ablation experiments. In addition, the results below are all calculated based on the effects of twenty iterations of these models. The proposed FFA's effectiveness is validated by adding it to CNN-Transformer and ResNet and using different parameters $p$ to illustrate the effect of FFA's weight matrix scales. The results of the ablation experiments are shown in \textbf{Table~\ref{ablation1}}, and to more obviously represent the impact of each module on the recognition performance, the amount of variation of metric was shown in \textbf{Table~\ref{ablation2}}. The absolute value of the variation bigger, the recognition performance better. We can observe that \textbf{i)} The average F1 score of CNN-Transformer with FFA is higher $0.9, 1.3, 2.0$ than the original one when the parameters $p=2, 4, 8$, respectively. Similarly, the ResNet with FFA compare to the original one has the same trend as CNN-Transformer. The improvements in the average F1 score are $2.35, 3.65, 7.2$, respectively. This means that the FFA can boost the model's recognition. The $p$ higher, the enhancement more significant. \textbf{ii)} When considering the effect of MMSA independently rather than the combination of FFA and MMSA, the recognition performance of CMT still outperforms CNN-Transformer (average F1 score is $4.5, 2.85, 1.2$ higher under $g/g'=2, 5, 10$). In addition, the smaller the MMSA parameters $g, g'$ (the finer the level of MMSA calculation), the better the CMT's performance. The superiority and effectiveness of the proposed MMSA over the conventional MHSA are validated. \textbf{iii)} Compared to the model without FFA and MMSA, the state-of-the-art (SOTA) performance was obtained for the whole CMT, and the best results were obtained for the $p=8$, $g=2, g'=2$ configuration, the average F1 score is $7.5$ higher than the conventional MHSA. All these imply that the proposed FFA can effectively eliminate the negative impact of the scarcity of medical image data and improve recognition accuracy. The absence of either the FFA or MMSA module negatively affects CMT, demonstrating their effectiveness and necessity.

\begin{table*}[htbp]
  \centering
  \caption{Results of the ablation experiments, which contain test results for different patch sizes ($p$) in FFA and various combinations of levels ($g, g'$) in the proposed MMSA. (\textbf{$\checkmark$}) means that the model carries a specific module (FFA or MMSA), while (\textbf{$\times$}) implies that the module does not exist. \textbf{(-)} represents the absence of this configuration, \textbf{(- / MMSA)} is only for CNN-Transformer that contains MMSA internally by default. The results for each class represent the recognition accuracy, with $\times 100$ for CP, CR, CF1, OP, OR, and OF1 results relative to the initial results.}
  \resizebox{\textwidth}{!}{
    \begin{tabular}{ccccccccccccccc}
    \toprule
    Model & FFA & FFA Configuration & MMSA & MMSA Configuration & COVID-19 & Normal & Lung Opacity & Viral pneumonia & CP & CR & CF1 & OP & OR & OF1 \\
    \midrule
    \midrule
    \multirow{4}[1]{*}{CNN-Transformer} & $\times$ & - & $\times$ & - / MHSA & 0.9806  & 0.8611  & 0.8992  & 0.9308  & 89.1  & 94.3  & 91.6  & 93.8  & 86.6  & 90.0 \\
        & $\checkmark$ & $p=2$ & $\times$ & - / MHSA & 0.9773  & 0.8820  & 0.9087  & 0.9410  & 92.1  & 93.8  & 92.9  & 93.9  & 87.5  & 90.6 \\
        & $\checkmark$ & $p=4$ & $\times$ & - / MHSA & 0.9800  & 0.8729  & 0.9324  & 0.9388  & 92.2  & 93.9  & 93.0  & 94.7  & 88.0  & 91.2 \\
        & $\checkmark$ & $p=8$ & $\times$ & - / MHSA & 0.9724  & 0.9027  & 0.9333  & 0.9489  & 92.7 & 92.8  & 92.8  & 94.6  & 90.1  & 92.8 \\
    \midrule
    \multirow{11}[1]{*}{CMT} & $\times$ & -  & $\checkmark$ & $g=2, g'=2$ & 0.9779  & 0.9340  & 0.9192  & 0.9181  & 94.1  & 96.9  & 95.5 & 94.2  & 96.1  & 95.1 \\
       & $\times$ & -  & $\checkmark$ & $g=5, g'=5$ & 0.9670  & 0.9028  & 0.8759  & 0.8907  & 91.1  & 96.8  & 93.9  & 90.6  & 96.5  & 93.4 \\
       & $\times$ & -  & $\checkmark$ & $g=10, g'=10$ & 0.9566  & 0.9227  & 0.9010  & 0.9102  & 92.5  & 91.3  & 92.1  & 93.6  & 90.3  & 91.9 \\
       & $\checkmark$ & $p=2$ & $\checkmark$ & $g=2, g'=2$ & 0.9836  & 0.9394  & 0.9245  & 0.9240  & 94.7  & 97.5  & 96.1  & 94.8  & 96.7  & 92.8 \\
       & $\checkmark$ & $p=4$  & $\checkmark$ & $g=2, g'=2$ & 0.9853  & 0.9486  & 0.9521  & 0.9191  & 94.1  & 97.9  & 96.0  & 95.8  & 96.4  & 96.1 \\
       & $\checkmark$ & $p=8$ & $\checkmark$ & $g=2, g'=2$ & 0.9972  & 0.9628  & 0.9714  & 0.9303  & 96.9  & 98.2  & 97.5  & 97.0  & 97.3  & 97.1 \\
       & $\checkmark$ & $p=2$ & $\checkmark$ & $g=5, g'=5$ & 0.9774  & 0.9032  & 0.8749  & 0.9560  & 92.6  & 95.4  & 94.0  & 91.7  & 95.3  & 93.5 \\
       & $\checkmark$ & $p=4$  & $\checkmark$ & $g=5, g'=5$ & 0.9842  & 0.9505  & 0.9370  & 0.9664  & 96.0  & 92.5  & 94.2  & 95.4  &  91.5  & 93.5 \\
       & $\checkmark$ & $p=8$ & $\checkmark$ & $g=5, g'=5$ & 0.9771  & 0.9568  & 0.9112  & 0.9802  & 95.6  & 92.8  & 94.3  & 94.8  & 93.4  & 94.1 \\
       & $\checkmark$ & $p=2$ & $\checkmark$ & $g=10, g'=10$ & 0.9760  & 0.9027  & 0.9293  & 0.9200  & 92.9  & 93.8  & 93.3  & 91.6  & 92.5  & 92.1 \\
       & $\checkmark$ & $p=4$  & $\checkmark$ & $g=10, g'=10$ & 0.9770  & 0.9157  & 0.9250  & 0.9163  & 93.4  & 94.3  & 93.8  & 92.8  & 94.1  & 93.5 \\
       & $\checkmark$ & $p=8$ & $\checkmark$ & $g=10, g'=10$ & 0.9841  & 0.9473  & 0.9311  & 0.9092  & 93.9  & 93.9  & 93.9  & 94.1  & 94.5  & 94.3 \\
    \midrule
    \multirow{4}[2]{*}{ResNet} 
       & $\times$ & - & $\times$ & -  & 0.9211 & 0.8366 & 0.9148 & 0.9018 & 82.1 & 89.4 & 85.6 & 87.2 & 82.3 & 84.6 \\
       & $\checkmark$ & $p=2$ & $\times$ & - & 0.9329  & 0.8555  & 0.9128  & 0.9315  & 86.5  & 91.2  & 88.8  & 87.4  & 84.9  & 86.1 \\
       & $\checkmark$ & $p=4$ & $\times$  & - & 0.9446  & 0.8791  & 0.9312  & 0.9284  & 89.5  & 90.4  & 89.9  & 88.3  & 87.0  & 87.6 \\
       & $\checkmark$  & $p=8$ & $\times$  & - & 0.9467  & 0.8678  & 0.9461  & 0.9015  & 90.0  & 94.5 & 92.6  & 94.3  & 89.9  & 92.0 \\
    \bottomrule
    \end{tabular}}
  \label{ablation1}%
\end{table*}%

\begin{table*}[htbp]
  \centering
  \caption{Amount of change in each evaluation metric relative to the initial model in the results of the ablation experiments. Each metric of the initial model is recorded as 0, (+) represents the amount of performance improvement relative to the initial model for the model trained in that configuration, and (-) represents the amount of performance degradation relative to the initial model for the model in that configuration.}
  \resizebox{1\textwidth}{!}{
    \begin{tabular}{cccccccccccc}
    \toprule
    Model & FFA & FFA Configuration & MMSA & MMSA Configuration & COVID-19 $\Delta$ & Normal $\Delta$ & Lung Opacity $\Delta$ & Viral Pneumonia $\Delta$ & Ave. $\Delta$ & CF1 $\Delta$ & OF1 $\Delta$ \\
    \midrule
    \midrule
    \multirow{4}[2]{*}{CNN-Transformer} & $\times$  & -  & $\times$  & MHSA & 0 & 0 & 0 & 0 & 0  & 0 & 0 \\
       & $\checkmark$  & $p=2$ & $\times$  & MHSA & -0.0033 & +0.0209 & +0.0095 & +0.0102 & +0.0094 & +1.3 & +0.6 \\
       & $\checkmark$  & $p=4$ & $\times$ & MHSA & -0.0006 & +0.0118 & +0.0332 & +0.0080 & +0.0130 & +1.4 & +1.2 \\
       & $\checkmark$  & $p=8$ & $\times$  & MHSA & -0.0082 & +0.0416 & +0.0341 & +0.0181 & +0.0214 & +1.2 & +2.8 \\
    \midrule
    \multirow{12}[2]{*}{CMT} & $\times$  & -  & $\times$  & MHSA & 0 & 0 & 0 & 0 & 0  & 0 & 0 \\
       & $\times$  & -  & $\checkmark$  & $g=2, g'=2$ & -0.0027 & +0.0729 & +0.0200 & -0.0127 & 0.0193 & +3.9 & +5.1 \\
       & $\times$  & -  & $\checkmark$  & $g=5, g'=5$ & -0.0136 & +0.0417 & -0.0233 & -0.0401 & -0.0083 & +2.3 & +3.4 \\
       & $\times$  & -  & $\checkmark$  & $g=10, g'=10$ & -0.0240 & +0.0616 & +0.0018 & -0.0206 & +0.0047 & +0.5 & +1.9 \\
       & $\checkmark$  & $p=2$ & $\checkmark$  & $g=2, g'=2$ & +0.0030 & +0.0783 & +0.0253 & -0.0068 & +0.0250 & +4.5 & +2.8 \\
       & $\checkmark$  & $p=4$ & $\checkmark$  & $g=2, g'=2$ & +0.0047 & +0.0875 & +0.0529 & -0.0117 & +0.0334 & +4.4 & +6.1 \\
       & $\checkmark$ & $p=8$ & $\checkmark$  & $g=2, g'=2$ & +0.0166 & +0.1017 & +0.0722 & -0.0005 & +0.0475 & +5.9 & +7.1 \\
       & $\checkmark$  & $p=2$ & $\checkmark$  & $g=5, g'=5$ & -0.0032 & +0.0421 & -0.0243 & +0.0252 & +0.0100 & +2.4 & +3.5 \\
       & $\checkmark$  & $p=4$ & $\checkmark$  & $g=5, g'=5$ & +0.0036 & +0.0894 & +0.0378 & +0.0356 & +0.0416 & +2.6 & +3.5 \\
       & $\checkmark$  & $p=8$ & $\checkmark$  & $g=5, g'=5$ & -0.0035 & +0.0957 & +0.0341 & +0.0494 & +0.0439 & +2.7 & +4.1 \\
       & $\checkmark$  & $p=2$ & $\checkmark$  & $g=10, g'=10$ & -0.0046 & +0.0416 & +0.0301 & -0.0108 & +0.0141 & +1.7 & +2.1 \\
       & $\checkmark$  & $p=4$ & $\checkmark$ & $g=10, g'=10$ & -0.0036 & +0.0546 & +0.0258 & -0.0145 & +0.0155 & +2.2 & +3.5 \\
       & $\checkmark$  & $p=8$ & $\checkmark$  & $g=10, g'=10$ & +0.0035 & +0.0862 & +0.0319 & -0.0216 & +0.0250 & +2.3 & +4.3 \\
    \midrule
    \multirow{4}[2]{*}{ResNet} & $\times$  & -  & $\times$  & -  & 0 & 0 & 0 & 0 & 0  & 0 & 0 \\
       & $\checkmark$  & $p=2$ & $\times$  & -  & +0.0118 & +0.0189 & -0.0020 & +0.0297 & +0.0146 & +3.2 & +1.5 \\
       & $\checkmark$  & $p=4$ & $\times$  & -  & +0.0235 & +0.0425 & +0.0164 & +0.0266 & +0.0273 & +4.3 & +3.0 \\
       & $\checkmark$  & $p=8$ & $\times$  & -  & +0.0256 & +0.0312 & +0.0313 & -0.0003 & +0.0220 & +7.0 & +7.4 \\
    \bottomrule
    \end{tabular}}
  \label{ablation2}%
\end{table*}%

\subsection{Interpretability Exploration}
\label{INTER}
\begin{figure}[tbh]
    \centering
    \includegraphics[width=0.9\textwidth]{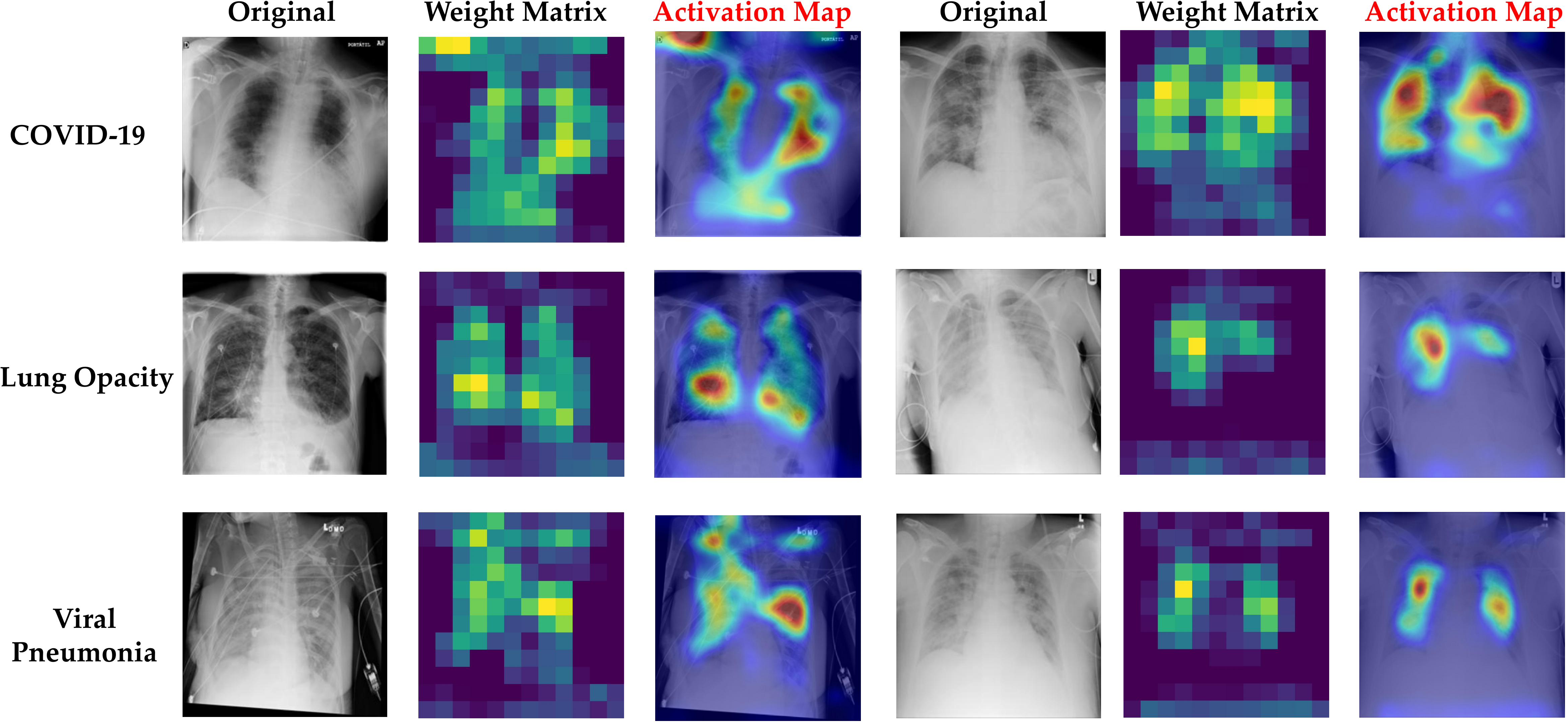}
    \caption{The representative weight matrices derived from the proposed CMT and their activation maps, from top to bottom, are the corresponding CXR images for COVID-19, Lung Opacity, and Viral pneumonia and their corresponding weights and activation maps.}
    \label{F7}
\end{figure}

DL models, long regarded as 'black boxes' due to their feature extraction and attention mechanism, are unclear and lack credibility in real medical applications. To explore the interpretability of the proposed CMT, we use Grad-CAM~\cite{selvaraju2017grad} to extract weights from the model in CXR image examples with lung disease and visualize the weighing matrix and feature activation maps for presenting the region of the model's attention. The visualization results are shown in Fig.~\ref{F7}. The region covered by darker color represents that the proposed CMT pays more attention to it and contains lots of task-related information. The proposed CMT focuses on the lung region, which has more influence on the recognition results than other regions. This demonstrates that our proposed CMT can focus more on the effective information across different regions in CXR images and ignore other noisy and minimally important features to obtain excellent recognition performance.

\section{Conclusion, Limitation, and Future work}
\label{CF}
In conclusion, we proposed a novel pneumonia recognition framework with interpretability and high efficiency for providing reliable, accurate analytic support to computer-aided medical clinic diagnosing systems. The proposed CNN-MMSA-Transformer (CMT) consists of CNN, considered a feature extractor, and a Transformer with a low computational complexity self-attention mechanism. Moreover, a low-cost CXR image data augmentation strategy has been introduced to reduce the negative implication on recognition accuracy because of the scarcity of medical image data. Experiments on the CXR image dataset fused from five distinct medical institutions have demonstrated the effectiveness of the proposed recognition framework. In addition, both the necessity and efficacy of the novel self-attention mechanism and data augmentation have been validated in further ablation studies.

Our work also has some limitations. First, as in most medical image recognition works that adopt the data-centric learning strategy, our works also run on the fusing CXR images dataset from different institutions. However, data interaction across various medical institutions is sometimes limited due to data exposure, and privacy concerns remain. For example, some hospital doesn't share their data with others due to confidentiality. Secondly, it still relies on the per-labeled data that requires extensive expert experience intervention to understand knowledge from CXR images during training. Finally, the proposed model still needs to consider all parameters' computations, although it is a high-performance and low computational complexity solution, which sometimes is unfriendly to satisfy the low-resourced constraint devices in hospitals.

In future work, aiming to solve the above three limitations, we will develop a general medical image recognition framework across medical institutions without direct data interaction to ensure data privacy. In the meantime, we will develop a new learning mechanism to obtain sufficiently superior recognition performance in understanding a few labeled CXR image datasets while reducing the model's parameter utilization.

\bibliography{main}
\bibliographystyle{tmlr}

\end{document}